\documentclass[journal]{IEEEtran}
\usepackage{algorithm}
\usepackage{algorithmic}
\IEEEoverridecommandlockouts
\usepackage{graphicx}
\usepackage{amsmath}
\usepackage{multirow}
\usepackage{multicol}
\usepackage{subfigure}
\usepackage{makecell}
\usepackage{bm}
\usepackage{lettrine}
\usepackage{amsfonts}
\newcommand{\PreserveBackslash}[1]{\let\temp=\\#1\let\\=\temp}
\newcolumntype{C}[1]{>{\PreserveBackslash\centering}p{#1}}
\newcolumntype{R}[1]{>{\PreserveBackslash\raggedleft}p{#1}}
\newcolumntype{L}[1]{>{\PreserveBackslash\raggedright}p{#1}}
\usepackage{threeparttable}

\usepackage{booktabs}
\usepackage{cite}
\usepackage[usenames]{color}
\usepackage{colortbl,booktabs}
\usepackage{tabularx}
\usepackage[Symbol]{upgreek}
\usepackage{balance}

\begin{document}

\bibliographystyle{IEEEtran} % use IEEEtran.bst style

% paper title
\title{Multiuser Diversity Gain from Superposition of Infinite Users over Block-Fading MAC}

\author{\IEEEauthorblockN{Yushu Zhang{$^{1}$}, Kewu Peng{$^{1,2}$},~\IEEEmembership{Senior Member,~IEEE}, Jian Song{$^{1,2}$},~\IEEEmembership{Fellow,~IEEE}, and Shuang Chen{$^{1}$}}

\IEEEauthorblockA{{$^{1}$}Electronic Engineering Department, Tsinghua University\\
Tsinghua National Laboratory for Information Science and Technology, Beijing 100084, P. R. China\\
{$^{2}$}Guangdong Province as well as Shenzhen City Key Laboratory of Digital TV System, Shenzhen 518057, P. R. China\\
Email:  pengkewu@tsinghua.edu.cn, zhang-ys15@mails.tsinghua.edu.cn}
%\author{\IEEEauthorblockN{Author~1, Author~2, Author~3, and Author~4}
%\author{\IEEEauthorblockN{Xu~Ma, Fang~Yang, \textsl{Senior Member, IEEE}, Sicong~Liu, \textsl{Student Member, IEEE}, Wenbo~Ding, \textsl{Student Member, IEEE}, and Jian~Song, \textsl{Fellow, IEEE}}
%\thanks{Author~1, Author~2, Author~3, and Author~4 are with Affiliation~1. Author~2 and Author~4 are also with Affiliation~2.}
%\thanks{This work was supported in part by the National Natural Science Foundation of China (Grant No. 61401248), in part by the New Generation Broadband Wireless Mobile Communication Network of the National Science and Technology Major Projects (Grant No. 2015ZX03002008), and in part by the R\&D Project of Science and Technology Innovation Commission of Shenzhen, China (No. GJHZ20130417162825486).}
%
%\thanks{Xu~Ma, Fang~Yang, Sicong~Liu, Wenbo~Ding, and Jian~Song are with the Electronic Engineering Department \& Research Institute of Information Technology, Tsinghua University, Tsinghua National Laboratory for Information Science and Technology (TNList), Beijing 100084, P. R. China. Email: \{maxu14, liu-sc12, dwb11\}@mails.tsinghua.edu.cn.
%
%Fang~Yang and Jian~Song are also with Shenzhen City Key Laboratory of Digital TV System, Shenzhen 518057, P. R. China. Email: \{fangyang, jsong\}@tsinghua.edu.cn.}
}
\maketitle
\begin{abstract}
In uplink block-fading multiple access channel (BF-MAC), the advantage of multiuser diversity (MUD) can be taken to achieve higher system throughput. In this letter, with rate constraints for all users and only channel state information available at the receiver assumed, we demonstrate that non-orthogonal multiple access (NOMA) outperforms the counterpart of orthogonal multiple access (OMA) via exploiting the MUD from the superposition of multiple users. The MUD gain achieved by NOMA compared with OMA is quantitatively analyzed for finite users, and the closed-form upper bound of MUD gain from the superposition of infinite users is also derived. Numerical results show that the potential MUD gain from superposition of infinite users can be well approached with limited superposed users, which indicates that multiple access schemes with limited superposed users can provide a good tradeoff between system performance and decoding complexity.

\end{abstract}


\IEEEpeerreviewmaketitle

\section{Introduction}
%Multiple access, widely believed as a key technology of the 5-th generation (5G) wireless communications to improve system throughput
Multiple access technology, widely believed as a key aspect of the 5-th generation (5G) wireless communications research to improve system throughput, has captured a lot of attentions recently~\cite{IMT2020}. In this letter, fading multiple access system, where multiple users transmit signals to one receiver, is considered. Due to independent fading characteristics of different users, the advantage of the inherent multiuser diversity (MUD) in received powers of different users can be taken by multiple access schemes to achieve higher system throughput~\cite{knopp1995information}.

In frequency-flat block-fading multiple access channel (BF-MAC), when perfect channel state information at the transmitters (CSIT) and the receiver (CSIR) is assumed, an optimal power control scheme is presented~\cite{knopp1995information} to exploit the MUD and maximize the throughput. In this scheme, only one user with the best channel is scheduled to transmit at one specific transmission block, and the throughput increases with the number of users due to MUD. Furthermore, the power control scheme is extended and the MUD in non-rich-scattering or slow-fading scenario is exploited via opportunistic beamforming~\cite{viswanath2002opportunistic}, where multiple transmit antennas are employed to induce large and fast channel fluctuations.

%multiple-input multiple-output (MIMO)
However, for delay-sensitive services, each user is required to transmit at a certain rate over a certain period, and maximizing the throughput is not the sole target. Thus, rate constraints for all users should be considered.

In this case, it is shown that non-orthogonal multiple access (NOMA), which allows all superposed users to share the whole channel resources simultaneously, can fully exploit the MUD and performs much better than orthogonal multiple access (OMA)~\cite{wang2006comparison}, which allocates channel resources exclusively to each user. With CSIT and CSIR assumed, NOMA can minimize the average-sum-power by allocating the transmission power for each user over each block on purpose. And for multiple-input multiple-output (MIMO) system, a quite simple and asymptotically optimal strategy called maximum eigen-mode beamforming is studied~\cite{wang2011maximum} for power minimization with rate constraints.

In this letter, we focus on exploiting the MUD for another typical scenario of delay-sensitive applications where instantaneous CSIT is not available. As a result, accurate power and rate control for each user over each block is not possible, and an outage event~\cite{li2005outage} will occur when the channel cannot support the transmission rates. Under these assumptions, it is found that NOMA can still exploit the MUD from the superposition of multiple users. Consequently, given the target transmission rates and target outage probabilities, lower transmission power is required by NOMA than OMA. This advantage of transmission power is called MUD gain in this letter.

This letter presents the MUD gain achieved by NOMA compared with OMA from the superposition of finite users over Rayleigh BF-MAC, and derives the closed-form upper bound of MUD gain with infinite superposed users. These results may provide an insight on the choice of the number of superposed users for future multiple access schemes to achieve a good tradeoff between system throughput and implementation complexity. This letter is organized as follows. Section II presents the system model. In section III, MUD gain achieved with finite superposed users is numerically analyzed, and the closed-form upper bound with infinite superposed users is derived. Finally, conclusions are drawn in Section IV.

\section{System Model}
Consider an uplink multiple access model, where $K$ users, $U=\{1,2,...,K\}$, transmit signals block-by-block to a single receiver with perfect CSIR and no CSIT. The signals of $K$ uplink users suffer from frequency-flat block fading, which indicates that the channel gain is frequency-flat and time-invariant during each received block, but changes independently from block to block.
For simplicity, in this letter, we mainly focus on symmetric users, having identical target transmission rate, ${R_s}/K$ (bps), and transmission power, ${P_s}/K$. Here, ${R_s}$  and ${P_s}$ denote the total transmission rate and the total transmission power, respectively. And the channel gains of $K$ users, $h_1,h_2,...,h_K$, are assumed to be independent and identically distributed (i.i.d.) Rayleigh fading. We refer to ${H_k} = {\left| {{h_k}} \right|^2}$ as channel parameter of User $k$, and with $\mathbb{E}\{ {{{\left| {{h_k}} \right|}^2}} \} = 1$ assumed, ${H_k}\sim\exp \left( 1 \right)$. $\mathbb{E}\{ \cdot\}$ denotes the expectation.

The total available bandwidth is $W$ (Hz) for a block, and is occupied by additive Gaussian white noise (AWGN) with spectral density of $N_0$. In the following discussion, we use the equivalent sum spectral efficiency ${\eta _s} = {R_s}/W$ (bps/Hz) and single-user spectral efficiency ${R_s}/(WK) = {\eta _s}/K$ (bps/Hz) instead of sum rate or single user rate.
%And the channel with bandwidth $W$ is partitioned into $K/J$ subchannels, each of which is shared by $J$ different users.
%The total available bandwidth is $W$ (Hz), occupied by additive Gaussian white noise with spectral density of $N_0$. The channel with bandwidth $W$ is partitioned into $K/J$ subchannels  and each subchannel is shared by $J$ different users. Note that each subchannel is equivalent to a MAC with $J$ uplink users when $J>1$. For simplicity, in this letter, we mainly focus on symmetric users, having identical target transmission rate, ${R_s}/K$ (bps), and transmission power, ${P_s}/K$. Here, ${R_s}$  and ${P_s}$ denote the total transmission rate and the total transmission power, respectively. Furthermore, the channel gains of K users are assumed to suffer from independent and identically distributed (i.i.d.) Rayleigh fading. $h_k$ denotes the channel gain of User $k$ over a block, and $\mathbb{E}| {{{\left| {{h_k}} \right|}^2}} | = 1$ is assumed for each user, where $\mathbb{E}|\cdot|$ denotes the expectation. In the following discussion, ${H_k} = {\left| {{h_k}} \right|^2}$ is referred to as channel parameter of User $k$, ${H_k}\sim\exp \left( 1 \right)$, and we use the equivalent sum spectral efficiency ${\eta _s} = {R_s}/W$ (bps/Hz) and single-user spectral efficiency (of User $k$) ${\eta _k} = {R_s}/(WK) = {\eta _s}/K$ (bps/Hz) instead of sum rate or single user rate.

\begin{figure}[t]
  \centering
  \includegraphics[width=0.49\textwidth]{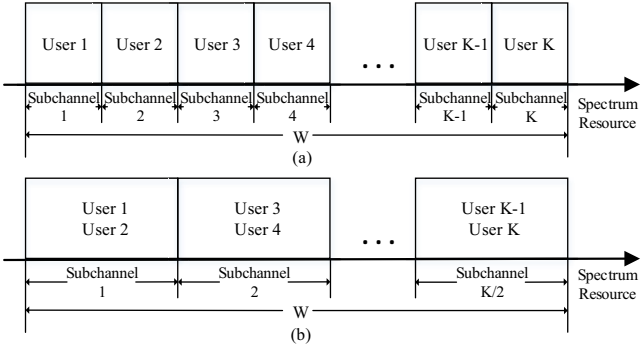}
  \caption{Resource allocation for (a) OMA, and (b) NOMA with $J=2$ superposed users.}\label{Fig1}
\end{figure}

For OMA, frequency-division multiple access (FDMA) is considered in this letter. As shown in Fig.~\ref{Fig1}(a), the total channel with bandwidth $W$ is partitioned into $K$ subchannels with bandwidth $W/K$, and each subchannel is used by only one user.

For NOMA with $J$ superposed users, the total bandwidth is partitioned into $K/J$ subchannels with bandwidth $WJ/K$, and each subchannel is shared by $J$ different users, where the total user number $K$ is assumed to be divisible by $J$ for simplicity. We assume a joint decoder is employed for NOMA. Under the assumption of symmetric users, the resource allocation for NOMA with $J=2$ superposed users is illustrated in Fig.~\ref{Fig1}(b).
% Note that each subchannel is equivalent to a MAC with $J$ uplink users when $J>1$.

The received signal-to-noise ratio (SNR) for User $k$ is
\begin{equation}\label{equ1}
{\gamma _k}\left( J \right) = \frac{{{H_k}{P_s}/K}}{{{N_0}JW/K}} = \frac{{{H_k}{P_s}}}{{{N_0}JW}},
\end{equation}
where $k=1,2,...,K$, and the SNR expression above includes OMA as a special case with $J=1$.

Due to the independent fading characteristics of different users, there exists inherent MUD in received powers from different users, and the MUD can be exploited by NOMA to improve sum spectral efficiency. Besides, the sum transmission power over unit spectrum could also increase with the number of superposed users, which is the inherent power gain in multiple access. In this letter, the sum transmission power over unit spectrum is fixed as $P_s/W$, and the effect of power gain can be excluded by controlling the total transmission power $P_s$ and the total bandwidth $W$. In this way, we can focus on the exploitation of MUD via NOMA.

%$\frac{{J{P_s}/K}}{{WJ/K}} = \frac{{{P_s}}}{W}
\section{MUD Gain from NOMA}
The improvement of system performance achieved by NOMA compared with that achieved by OMA is referred to as MUD gain in this letter. There are three equivalent forms to express MUD gain:

\hangindent 1.1em
1. The decrease in transmission power given target spectral efficiency and outage probability;

\hangindent 1.1em
2. The increase in spectral efficiency given target transmission power and outage probability;

\hangindent 1.1em
3. The decrease in outage probability given target transmission power and spectral efficiency.

This letter analyzes the MUD gain mainly from the perspective of the 1st form above.
\subsection{MUD Gain with Finite Superposed Users}
Given the sum spectral efficiency ${\eta _s}$ and target individual outage probability ${\varepsilon _{\textup{Ind}}}$ for $K$ active users, the required sum transmission power for a multiple access technique is denoted by ${P_s}\left( {{\eta _s},{\varepsilon _{\textup{Ind}}},J} \right)$, where $J=1$ represents OMA and $J>1$ represents NOMA with $J$ superposed users.

For OMA, an individual outage occurs to User $k$ when its channel gain cannot support the target single-user spectral efficiency, i.e. channel parameter $H_k$ satisfies
\begin{equation}\label{equ2}
\frac{{{\eta _s}}}{K} > \frac{1}{K}{\log _2}\left( {1 + \frac{{{H_k}{P_s}/K}}{{{N_0}W/K}}} \right),
\end{equation}
or equivalently,
\begin{equation}\label{equ3}
{H_k} < \frac{{{N_0}W}}{{{P_s}}}\left( {{2^{{\eta _s}}} - 1} \right).
\end{equation}

In this way, the individual outage probability ${\varepsilon _{\textup{Ind}}}$ for OMA can be calculated

\begin{eqnarray}\label{equ4}
\varepsilon _{\textup{Ind}}} = \Pr \left[ {{H_k} < \frac{{{N_0}W}}{{{P_s}}}\left( {{2^{{\eta _s}}} - 1} \right)} \right] = 1 - {e^{ - \frac{{{N_0}W}}{{{P_s}}}\left( {{2^{{\eta _s}}} - 1} \right)},
\end{eqnarray}
and the expression for ${P_s}\left( {{\eta _s},{\varepsilon _{\textup{Ind}}},J=1} \right)$ is obtained
\begin{equation}\label{equ5}
{P_s}\left( {{\eta _s},{\varepsilon _{\textup{Ind}}},1} \right) =  - \frac{{{N_0}W({2^{{\eta _s}}} - 1)}}{{\ln (1 - {\varepsilon _{\textup{Ind}}})}}.
\end{equation}

For NOMA with $J$ superposed users, if the vector of target spectral efficiencies of $J$ superposed users lies outside the achievable individual outage rate region of User $k$ over a given block~\cite{li2005outage}, an individual outage occurs to User $k$, which implies User $k$ cannot be decoded by the joint decoder regardless of the success or failure of the other $J\!-\!1$ users over this block. And via Monte-Carlo simulation, the required sum transmission power ${P_s}\left( {{\eta _s},{\varepsilon _{\textup{Ind}}},J} \right)$ for NOMA can be obtained.

\begin{figure}[t]
  \centering
  \includegraphics[width=0.49\textwidth]{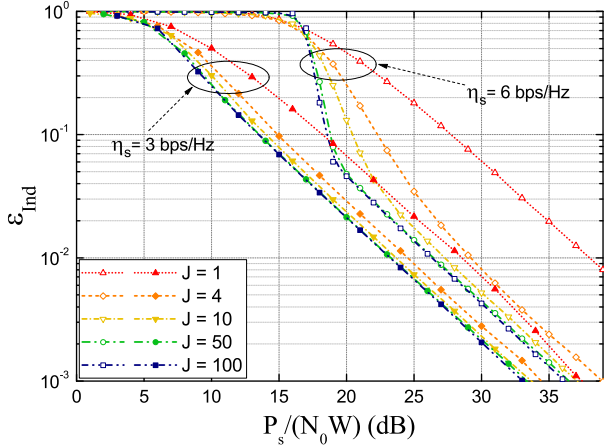}
  \caption{Individual outage probability vs. $P_s/(N_0W)$ with $\eta_s=3$ bps/Hz and $\eta_s=6$ bps/Hz.}\label{Fig2}
\end{figure}
We assume the total AWGN power, i.e. ${N_0}W$, is normalized. Via Monte-Carlo simulation, the individual outage probabilities for different sum transmission powers are obtained. As shown in Fig.~\ref{Fig2}, with target sum spectral efficiency $\eta_s=3$ bps/Hz, the required sum transmission power of NOMA with $J$ superposed users is lower than that of OMA, and decreases as $J$ becomes larger. Meanwhile, the gap between curves of $J=50$ and $J=100$ is very small at $\varepsilon_{\textup{Ind}}=0.01$, which implies the existence of a lower bound of individual outage probability. The same pattern holds for numerical results with $\eta_s=6$ bps/Hz, and the gap between curves of NOMA with $J$ superposed users and OMA at $\varepsilon_{\textup{Ind}}=0.01$ are even larger than that in the case of $\eta_s=3$ bps/Hz.

\begin{figure}[t]
  \centering
  \includegraphics[width=0.49\textwidth]{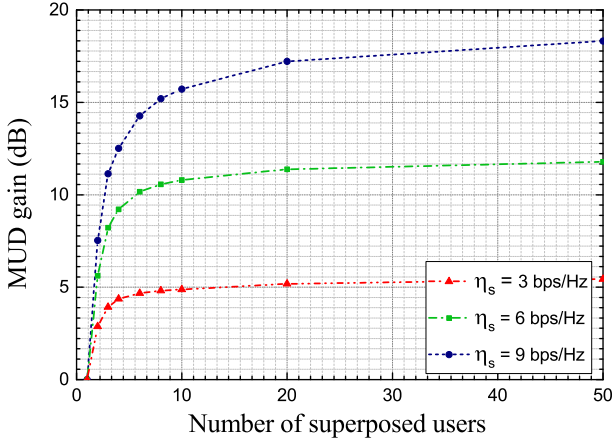}
  \caption{MUD gain vs. Number of superposed users with ${\varepsilon _{\textup{Ind}}} = 0.01$.}\label{Fig3}
\end{figure}
The power gain achieved by NOMA with $J$ superposed users compared with OMA is defined as MUD gain (in dB form), which can be expressed as follow
\begin{equation}\label{equ6}
{G_{\textup{MUD}}}\left( {{\eta _s},{\varepsilon _{\textup{Ind}}},J} \right) {\buildrel \Delta \over =} {P\!_s}\!\left( {{\eta _s},{\varepsilon _{\textup{Ind}}},1} \right)(\!\textup{dB}\!) - {P\!_s}\!\left( {{\eta _s},{\varepsilon _{\textup{Ind}}},J} \right)(\!\textup{dB}\!).
\end{equation}

The numerical results for MUD gain with finite superposed users at ${\varepsilon _{\textup{Ind}}} = 0.01$ are depicted in Fig.~\ref{Fig3}. The MUD gain is larger with higher sum spectral efficiency for a certain number of superposed users. And given the sum spectral efficiency, the MUD gain increases with the number of superposed users $J$, and the increase is rapid for small $J$ values and quite slow for large $J$ values. Taking $\eta_s=6$ bps/Hz as an example, the available MUD gain is 9.21 dB for $J=4$, and only increases to 11.78 dB for $J=50$. This simulates the motivation to investigate the upper bound of MUD gain as $J$ approaches infinity, which will be analyzed in the next subsection.
\subsection{The Upper Bound of MUD Gain with Infinite Superposed Users}
In this subsection, with infinite superposed users, the lower bound of individual outage probability and sum transmission power are firstly obtained, and then we derive the closed-form upper bound of MUD gain. To achieve maximum MUD gain, all active users should share the total bandwidth, i.e. $J=K$, and a sufficiently large $K$ is assumed. For a given block, let ${U_{\textup{IndOut}}}$ represent the set of all users with individual outage.
%all active users can be grouped into two sets, one set contains decodable users, and the other set ${U_{IndOut}}$ contains users with individual outage.
% The signals of users in ${U_{Suc}}$ can be decoded successfully and removed from the received signals, and the decoding process of users in ${U_{IndOut}}$ will still fail even without the interferences from users in ${U_{Suc}}$.

First, given sum transmission power $P_s$ and sum spectral efficiency $\eta_s$, the lower bound of individual outage probability is considered. If ${U_{\textup{IndOut}}}$ is not empty, there always exist a set of users, denoted by ${\tilde U_{\textup{IndOut}}}$, whose received SNRs are too low to support the target single-user spectral efficiency ${\eta _s}/K$ even though there is no interference from any other user in $U$. Besides, if ${U_{\textup{IndOut}}}=\o$, ${\tilde U_{\textup{IndOut}}}=\o$. The received SNR ${\gamma _{k'}}\left( J \right)$ of User $k'\in{\tilde U_{\textup{IndOut}}}$ satisfies
\begin{equation}\label{equ7}
\frac{{{\eta _s}}}{K} > {\log _2}\left( {1 + {\gamma _{k'}}\left( {J = K} \right)} \right),
\end{equation}
using (\ref{equ1}), inequation (\ref{equ7}) can be expressed as
\begin{equation}\label{equ8}
{H_{k'}} < \frac{{{N_0}WK}}{{{P_s}}}\left( {{2^{\frac{{{\eta _s}}}{K}}} - 1} \right).
\end{equation}
And given the assumption of symmetric users, for any User $k \in U$, $k \in {\tilde U_{\textup{IndOut}}}$ if and only if $H_k$ satisfies (\ref{equ8}).

%The individual outage probability can be expressed as  , where   represents the cardinality of  .
Since any user in ${\tilde U_{\textup{IndOut}}}$ cannot be successfully decoded even if there are no interferences, we have ${\tilde U_{\textup{IndOut}}} \subseteq {U_{\textup{IndOut}}}$, and received SNRs of other users in ${{U_{\textup{IndOut}}}\!-\!\tilde U_{\textup{IndOut}}}$ are larger than those of users in ${\tilde U_{\textup{IndOut}}}$ but will still fail due to multiuser interference. Using (\ref{equ8}) we can obtain a lower bound of individual outage probability $\varepsilon _{\textup{Ind}}^{\textup{LB}}$ as
\begin{eqnarray}
\label{equ9_1}{\varepsilon _{\textup{Ind}}}\! =\! \frac{{\left| {{U_{{\textup{IndOut}}}}} \right|}}{K}\!\!\!\!\! &\ge&\!\!\!\!\! \frac{{| {{{\tilde U}_{{\textup{IndOut}}}}} |}}{K}\\
 \label{equ9_2}\!\!\!\!&\buildrel \Delta \over =&\!\!\!\! \varepsilon _{\textup{Ind}}^{\textup{LB}}\! = \Pr\!\!\left[ {{H_{k'}} \!<\! \frac{{\!{N_0}\!W\!K\!}}{{{P_s}}}\!\!\left( \!{{2^{\frac{{{\eta _s}}}{K}}}\!\! - \!1}\! \right)} \!\right]\!\!,
\end{eqnarray}
where $|\cdot|$ denotes the cardinality of a set. Note that $\varepsilon _{\textup{Ind}}^{\textup{LB}} $ will be pretty close to ${\varepsilon _{\textup{Ind}}}$ when ${\varepsilon _{\textup{Ind}}}$ is low, because $|{{U_{\textup{IndOut}}}-\tilde U_{\textup{IndOut}}}|$ becomes smaller as ${\varepsilon _{\textup{Ind}}}$ becomes lower. Let $x=1/K$ and $f(x) \buildrel \Delta \over = \left( {{2^{{\eta _s}x}} - 1} \right){{N_0}W}/({P_s}x)$, representing the right side of (\ref{equ8}). The Taylor series expansion for $f(x)$ at the point $x=0$ is given as
\begin{eqnarray}\label{equ10}
\nonumber f(x) &=& \frac{{{N_0}W}}{{{P_s}x}}\left( {1 + ({\eta _s}\ln 2)x + {{\left( {{\eta _s}\ln 2} \right)}^2}\frac{{{x^2}}}{2} + o\left( {{x^2}} \right) - 1} \right)\\
 &=& \frac{{{N_0}W{\eta _s}\ln 2}}{{{P_s}}} + \frac{{{N_0}W}}{{{P_s}}}{\left( {{\eta _s}\ln 2} \right)^2}\frac{x}{2} + o\left( x \right).
\end{eqnarray}

Now we let $K$ tend to infinity. Since the channel parameters of different users are assumed to be i.i.d. and follow exponential distribution with mean of 1, the channel parameters of all active users over a given block follow the same distribution. By substituting (\ref{equ10}) in (\ref{equ9_2}), $\varepsilon _{\textup{Ind}}^{\textup{LB}}$ can be expressed as follow
\begin{eqnarray}\label{equ11}
\nonumber\varepsilon _{\textup{Ind}}^{\textup{LB}} \!\!\!&=&\!\!\! \mathop {\lim }\limits_{K \to \infty } \Pr \bigg[ {{H_{k'}} \!<\! \frac{{{N_0}WK}}{{{P_s}}}\left( {{2^{\frac{{{\eta _s}}}{K}}} - 1} \right)} \bigg]\\
\nonumber &=&\!\!\! \Pr \left[ {{H_{k'}} \!<\! \mathop {\lim }\limits_{K \to \infty } \frac{{{N_0}WK}}{{{P_s}}}\left( {{2^{\frac{{{\eta _s}}}{K}}} - 1} \right)} \right]\\
\nonumber &=&\!\!\! \Pr \bigg[ {{H_{k'}}\! <\! \mathop {\lim }\limits_{x \to 0} f(x)} \bigg]=\Pr \left[ {{H_{k'}} \!< \!\frac{{{N_0}W{\eta _s}\ln 2}}{{{P_s}}}} \right]\\
 \!&=&\!\!\! 1 - {2^{ - {N_0}W{\eta _s}/{P_s}}}.
\end{eqnarray}
\begin{figure}[t]
  \centering
  \includegraphics[width=0.50\textwidth]{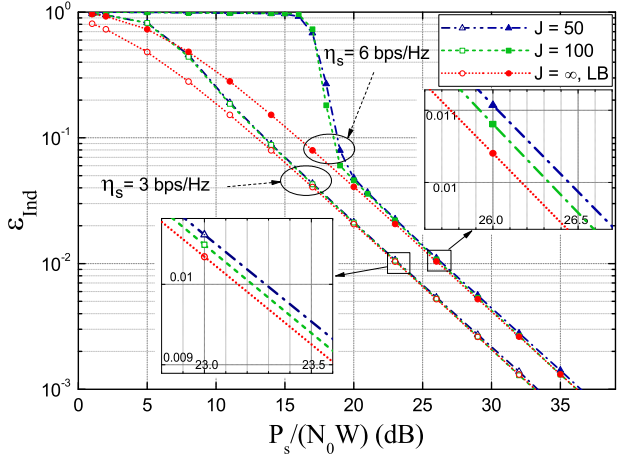}
  \caption{Lower bound of individual outage probability for NOMA with infinite superposed users, and individual outage probability for NOMA with finite superposed users for comparison.}\label{Fig4}
\end{figure}

The lower bound of individual outage probability $\varepsilon _{\textup{Ind}}^{\textup{LB}}$ with $J=\infty$, as well as the numerical results of NOMA with $J=50$ and $J=100$, are depicted in Fig.~\ref{Fig4}. And as discussed in the derivation above, the gap between curves of $J=\infty$ and $J=100$ is very small at relatively low individual outage probability for both cases of $\eta_s=3$ bps/Hz and $\eta_s=6$ bps/Hz.

In this way, the lower bound of required sum transmission power for NOMA $P_{s,{\textup{NOMA}}}^{\textup{LB}}$, given individual outage probability ${\varepsilon _{\textup{Ind}}}$ and total spectral efficiency $\eta_s$, can be obtained by substituting (\ref{equ11}) into (\ref{equ9_1})
\begin{equation}\label{equ12}
{P_s} \ge   - \frac{{{\eta _s}{N_0}W}}{{{{\log }_2}(1 - {\varepsilon _{\textup{Ind}}})}} \buildrel \Delta \over = P_{s,{\textup {NOMA}}}^{\textup{LB}}.
\end{equation}
After normalizing the total AWGN power $N_0W$ to unity, and according to (\ref{equ5}), (\ref{equ6}) and (\ref{equ12}), the upper bound of MUD gain (in dB form) is derived as
%\begin{equation}\label{equ13}
%\resizebox{1\hsize}{!}{$ G_{\textup{M\!U\!D}}^{\textup{U\!B}}\!\!\left( {{\eta_s},{\varepsilon _{\textup{\!Ind}}}} \right) \!\buildrel \Delta \over =\! 10{\log _{10}}\!\!\left(\!\! { - \frac{{({2^{{\eta\!_s}}}\! - 1)}}{{\ln (1 \!-\! {\varepsilon _{\textup{\!Ind}}})}}} \!\!\right) - 10{\log _{10}}\!\!\left(\! \!{ - \frac{{{\eta_s}}}{{{{\log }_2}\!(1\! -\! {\varepsilon _{\textup{\!Ind}}})}}} \!\!\right).$}
%\end{equation}
\begin{eqnarray}\label{equ13}
\nonumber G_{{\textup{MUD}}}^{{\textup{UB}}}\left( {\eta {_s},{\varepsilon _{{\textup{Ind}}}}} \right) \buildrel \Delta \over = {P_s}\left( {{\eta _s},{\varepsilon _{{\textup{Ind}}}},1} \right)(\textup{dB}) - P_{s,{\textup{NOMA}}}^{{\textup{LB}}}(\textup{dB})\\
= 10{\log _{10}}\!\!\left(\!\! { - \frac{{({2^{{\eta_s}}}\! - 1)}}{{\ln (1 \!-\! {\varepsilon _{\textup{Ind}}})}}} \!\!\right) \!-\! 10{\log _{10}}\!\!\left(\! \!{ - \frac{{{\eta_s}}}{{{{\log }_2}\!(1\! -\! {\varepsilon _{\textup{Ind}}})}}} \!\!\right).
\end{eqnarray}

\begin{figure}[t]
  \centering
  \includegraphics[width=0.50\textwidth]{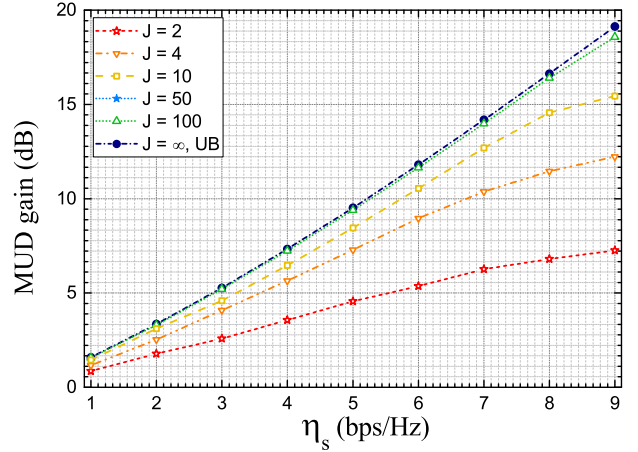}
  \caption{MUD gain vs. sum spectral efficiency, with ${{\varepsilon _{\textup{Ind}}}=0.01}$.}\label{Fig5}
\end{figure}
In Fig.~\ref{Fig5}, the upper bound of MUD gain derived in (\ref{equ13}) is presented, and several curves of MUD gain achieved by NOMA with finite users are also obtained for comparison. The gap between the curves of $J=100$ and the upper bound of MUD gain with $J=\infty$ is very small at ${{\varepsilon _{\textup{Ind}}}=0.01}$ from $\eta_s=1$ bps/Hz to $\eta_s=9$ bps/Hz. Moreover, given sum spectral efficiency, the increase of MUD gain becomes slower as $J$ becomes larger. At ${\eta _s} = 3$ bps/Hz, NOMA schemes with only 2, 4, and 10 superposed users achieve 49.0$\%$, 77.4$\%$ and 87.0$\%$ of the upper bound of MUD gain $ G_{\textup{MUD}}^{\textup{UB}}$, respectively, and at ${\eta _s} = 6$ bps/Hz, the corresponding portions of $ G_{\textup{MUD}}^{\textup{UB}}$ achieved by NOMA schemes with 2, 4 and 10 superposed users are 45.4$\%$, 75.9$\%$ and 89.3$\%$, respectively. A major portion of potential MUD gain is available for NOMA with limited superposed users, and since the decoding complexity increases with $J$, NOMA schemes with limited superposed users could achieve a good tradeoff between system performance and decoding complexity.
%Since a major portion of potential MUD gain is available to NOMA with limited superposed users,  could achieve,

\section{Conclusions}
In this letter, we exploit the MUD via NOMA over Rayleigh BF-MAC, under the assumption of rate constraints and no instantaneous CSIT. Firstly, the MUD gain achieved by NOMA with finite superposed users is quantitatively analyzed compared with OMA. Then the closed-form upper bound of MUD gain with infinite superposed users is derived. Numerical results show that MUD gain achieved by NOMA increases with higher sum spectral efficiency and more superposed users. Moreover, NOMA with only 4 superposed users could achieve at least 77.4$\%$ and 75.9$\%$ of the potential MUD gain with infinite superposed users with sum spectral efficiency of 3 bps/Hz and 6 bps/Hz, respectively, at individual outage probability of 0.01. This indicates that multiple access schemes with limited superposed users could achieve a considerable portion of potential MUD gain, and provide a good tradeoff between system performance and decoding complexity.

\bibliography{IEEEabrv,bibfile}

\end{document}